\documentclass[twoside,twocolumn,9pt]{article}
\usepackage[utf8]{inputenc}
\usepackage{extsizes}
\usepackage[super,sort&compress,comma]{natbib} 
\usepackage[version=3]{mhchem}
\usepackage[left=1.5cm, right=1.5cm, top=1.785cm, bottom=2.0cm]{geometry}
\usepackage{balance}
\usepackage{times,mathptmx}
\usepackage{sectsty}
\usepackage{graphicx} 
\usepackage{lastpage}
\usepackage[format=plain,justification=justified,singlelinecheck=false,font={stretch=1.125,small,sf},labelfont=bf,labelsep=space]{caption}
\usepackage{float}
\usepackage{fancyhdr}
\usepackage{fnpos}
\usepackage[english]{babel}
\addto{\captionsenglish}{%
  
}
\usepackage{amsmath}
\usepackage{amssymb}
\DeclareMathOperator{\Tr}{Tr}
\usepackage{array}
\usepackage{droidsans}
\usepackage{charter}
\usepackage[T1]{fontenc}
\usepackage[usenames,dvipsnames]{xcolor}
\usepackage{setspace}
\usepackage[compact]{titlesec}
\usepackage{hyperref}
\usepackage{epstopdf}

\definecolor{cream}{RGB}{222,217,201}

\begin{document}

\pagestyle{fancy}
\thispagestyle{plain}
\fancypagestyle{plain}{

\renewcommand{\headrulewidth}{0pt}
}

\makeFNbottom
\makeatletter
\renewcommand\LARGE{\@setfontsize\LARGE{15pt}{17}}
\renewcommand\Large{\@setfontsize\Large{12pt}{14}}
\renewcommand\large{\@setfontsize\large{10pt}{12}}
\renewcommand\footnotesize{\@setfontsize\footnotesize{7pt}{10}}
\makeatother

\renewcommand{\thefootnote}{\fnsymbol{footnote}}
\renewcommand\footnoterule{\vspace*{1pt}%
\color{cream}\hrule width 3.5in height 0.4pt \color{black}\vspace*{5pt}} 
\setcounter{secnumdepth}{5}

\makeatletter 
\renewcommand\@biblabel[1]{#1}            
\renewcommand\@makefntext[1]%
{\noindent\makebox[0pt][r]{\@thefnmark\,}#1}
\makeatother 
\renewcommand{\figurename}{\small{Fig.}~}
\sectionfont{\sffamily\Large}
\subsectionfont{\normalsize}
\subsubsectionfont{\bf}
\setstretch{1.125} 
\setlength{\skip\footins}{0.8cm}
\setlength{\footnotesep}{0.25cm}
\setlength{\jot}{10pt}
\titlespacing*{\section}{0pt}{4pt}{4pt}
\titlespacing*{\subsection}{0pt}{15pt}{1pt}

\fancyfoot{}
\fancyfoot[RO]{\footnotesize{\sffamily{1--\pageref{LastPage} ~\textbar  \hspace{2pt}\thepage}}}
\fancyfoot[LE]{\footnotesize{\sffamily{\thepage~\textbar\hspace{3.45cm} 1--\pageref{LastPage}}}}
\fancyhead{}
\renewcommand{\headrulewidth}{0pt} 
\renewcommand{\headrulewidth}{0pt}
\setlength{\arrayrulewidth}{1pt}
\setlength{\columnsep}{6.5mm}
\setlength\bibsep{1pt}

\makeatletter 
\newlength{\figrulesep} 
\setlength{\figrulesep}{0.5\textfloatsep} 

\newcommand{\topfigrule}{\vspace*{-1pt}%
\noindent{\color{cream}\rule[-\figrulesep]{\columnwidth}{1.5pt}} }

\newcommand{\botfigrule}{\vspace*{-2pt}%
\noindent{\color{cream}\rule[\figrulesep]{\columnwidth}{1.5pt}} }

\newcommand{\dblfigrule}{\vspace*{-1pt}%
\noindent{\color{cream}\rule[-\figrulesep]{\textwidth}{1.5pt}} }

\makeatother


\twocolumn[
  \begin{@twocolumnfalse}
\vspace{3cm}
\sffamily
\begin{tabular}{m{4.5cm} p{13.5cm} }

 &\noindent\LARGE{\textbf{Dispersion of activity at an active-passive nematic interface}} \\
\vspace{0.3cm} & \vspace{0.3cm} \\

 & \noindent\large{Rodrigo C. V. Coelho,$^{\ast}$\textit{$^{a,b}$} Nuno A. M. Araújo,\textit{$^{a,b}$} and Margarida M. Telo da Gama\textit{$^{a,b}$}} \\\\

&\noindent\normalsize{
Efficient nutrient mixing is crucial for the survival of bacterial colonies and other living systems known as active nematics. However, the dynamics of this mixing is non-trivial as there is a coupling between nutrients concentration and velocity field. To address this question, we solve the hydrodynamic equation for active nematics to model the bacterial swarms coupled to an advection-diffusion equation for the activity field, which is proportional to the concentration of nutrients. At the interface between active and passive nematics the activity field is transported by the interfacial flows and in turn it modifies them through the generation of active stresses. We find that the dispersion of this conserved activity field is subdiffusive due to the emergence of a barrier of negative defects at the active-passive interface, which hinders the propagation of the motile positive defects. 
} 

\\

\end{tabular}

 \end{@twocolumnfalse} \vspace{0.6cm}

  ]

\renewcommand*\rmdefault{bch}\normalfont\upshape
\rmfamily
\section*{}
\vspace{-1cm}


\footnotetext{\textit{$^{a}$~Centro de Física Teórica e Computacional, Universidade de Lisboa, 1749-016 Lisboa, Portugal; E-mail: rcvcoelho@fc.ul.pt}}
\footnotetext{\textit{$^{b}$~Departamento de Física, Faculdade de Ciências,
Universidade de Lisboa, P-1749-016 Lisboa, Portugal.}}





\section{Introduction}

Active matter is made of particles that convert energy into mechanical motion. The size of the particles varies over a wide range of scales, from macroscopic birds in flocks down to the sub-cellular scale of microtubules~\cite{MarchettiRMP2013,doi:070909-104101,VICSEK201271,RevModPhys.88.045006}. In many active systems, the particles are elongated and exhibit local nematic order. In wet systems the particles are suspended in fluids and momentum conservation ensues. Active wet nematics are characterized by instabilities that drive the uniform nematic state into an active turbulent state at any level of activity~\cite{Sanchez2012} and the theory of active nematics has played a role in establishing a framework to describe this class of active systems~\cite{PhysRevLett.89.058101}. 

Examples of wet active matter with elongated particles are mixtures of microtubule-kinesin and colonies of bacteria. In mixtures of microtubule-kinesin the presence of half-integer point defects in 2D and disclination loops in 3D, as well as their complex dynamics, were observed experimentally~\cite{Sanchez2012, doi:10.1126/science.aaz4547}. The local nematic order and the activity driven dynamics are described well by the active nematics hydrodynamic model. Experimental evidence for active nematics behaviour in bacterial swarms, namely the presence and dynamics of topological defects, is less clear. 
Recently, however, the phenomenology of wet active nematics was reported in colonies of Serratia marcescens bacteria, which were allowed to grow and elongate forming a quasi 2D system of particles with local nematic order~\cite{Li777}. The dynamics of the quasi point-like defects was found to be in line with that of active nematics, although details of the motion and structure of the defects revealed that some of the assumptions of 2D active nematics are oversimplified and need to be relaxed to account for the observed features. In dry systems, active nematics behaviour was also reported as signalled by the presence of mobile defects and tangential interfacial anchoring in growing 2D colonies of rod-shaped E. coli~\cite{DellArciprete2018}.

In wet active systems, the nutrients that provide the source of energy are, often, heterogeneously dispersed and need to be distributed throughout the system to sustain its activity. For instance, in microtubule-kinesin, the activity is controlled by the concentration of ATP~\cite{doi:10.1098/rsta.2014.0142, Hardoin2019, PhysRevLett.127.148001} and this concentration can be heterogeneous. 

In the active turbulent state the nutrients and particles are mixed even at very low Reynolds numbers. Active turbulence arises in wet and non-wet active systems and resembles inertial turbulence at high Reynolds numbers~\cite{ PhysRevLett.110.228102,alert082321-035957, Alert2020}. This flow state, characterized by the statistical properties of its energy spectrum, has been reported for a wide range of active systems including swarming bacteria~\cite{Wensink14308, Patteson2018}. Are there any indications that these biological systems evolved to optimize nutrient mixing? 

Recently~\cite{C9SM02306B}, we reproduced the statistical properties of Serratia marcescens bacterial swarms using the hydrodynamic model of active nematics subject to substrate friction. Most studies (including ours)~\cite{alert082321-035957} consider a homogeneous static activity field, which may be a strong assumption for many cases of interest. The activity field and its spatial and time evolution could be relevant to understanding nutrient mixing at low Reynolds numbers. The mechanism of mixing in active nematics was described in Ref.~\cite{Tan_2019}. A related question is how does a heterogeneous distribution of nutrients evolve in wet and non-wet active systems? This question has received little or no attention. 

Related experimental studies of the dispersion of inert particles (tracers) in a bacterial bath, in the active turbulent state, revealed the emergence of coherent structures, swirls and jets~\cite{PhysRevLett.84.3017, Kurtuldu10391}. At short times, the motion of the tracers is superdiffusive and it becomes diffusive at longer times. Furthermore, the diffusion of the tracers increases with the concentration of bacteria, by contrast to the behaviour of passive systems. Another interesting observation is that the average vorticity in the bacterial bath is maximal at a specific aspect ratio, decreasing for more elongated bacteria~\cite{Beer2019}. Since vorticity is related to the efficiency of solute mixing, this observation suggests that there may be an optimal shape for nutrient mixing in active turbulent systems. How are tracers dispersed in active turbulent nematics where the activity is heterogeneous? Can this be used to measure the evolution of activity in systems with a heterogeneous distribution of nutrients? 

Heterogeneous static activity fields have been investigated recently~\cite{Zhang2021, PhysRevLett.126.227801, Zhang0.1038/s41578} both theoretically and experimentally. In particular, static spatial patterns of activity were used to generate defects and control their motion within the active region. In addition, theoretical investigations of the distribution of defects and their orientation at static active-passive interfaces were reported in Refs.~\cite{PhysRevX.9.041047, PhysRevE.103.022703}. In all cases the boundaries between the active and the passive domains are fixed and the activity field is bound to the active region. How is the structure of these static active-passive interfaces related to that of dynamical (propagating) interfaces in systems where the activity evolves in space and time?     

In addition to the question of nutrient mixing, the spatial and temporal evolution of a heterogeneous distribution of activity and of the corresponding propagating active-passive interface is also relevant in the context of dispersion of antibiotics in biofilms~\cite{10.1016/j.jpha.2015.11.005}. 

In what follows, we investigate the dispersion of a scalar field that generates active stresses in model active nematics. We couple a hydrodynamic model for active nematics (e.g., used in Refs.~\cite{C9SM00859D, C9SM02306B}) with an advection-diffusion equation for the concentration of the scalar activity field. The conserved activity spreads from the active to the passive region by diffusion and by the flow, which in turn is driven by the active stresses. As a result the active-passive interface propagates throughout the system, exhibiting a novel subdiffusive regime.
This subdifusive regime results from the polarization of the moving active-passive interface due to the emergence of a barrier of nearly static negative defects, which hinders the propagation of the motile positive ones. Finally, investigate the dependence of the generalized diffusion coefficient on the activity and on the aligning parameter. We find a non-monotonic dependence on the aligning parameter, suggesting the existence of an optimal shape that maximizes the dispersion of the nutrients.

The paper is organized as follows. In Sec.~\ref{method-sec}, we describe the model used to simulate the dispersion of the activity field. In Sec.~\ref{sharp-sec}, we analyse the motion of the defects in simple geometries characterized by initially sharp active-passive interfaces. 
In Sec.~\ref{dispersion-sec}, the dispersion of activity is studied for more realistic initially smooth (Gaussian) activity fields as a function of the aligning parameter. For comparison we also study the dynamics of inert tracers, which by contrast to the activity field do not produce active stresses. Finally, in Sec.~\ref{conclusions-sec}, we summarize our findings.

\section{Method}
\label{method-sec}

Dispersion is the process of spreading the concentration of a solute where the flux depends both on the concentration gradient and on the fluid velocity. Diffusion is a special case of dispersion when the fluid velocity is zero while advection is the transport of the solute by the fluid velocity without changing its concentration in the comoving frame.
In order to simulate the dispersion of a conserved activity field (e.g., nutrients or ATP) by the chaotic flow of an active turbulent nematic, we couple the hydrodynamic model of active nematics used in previous works (for instance, Refs.~\cite{C9SM00859D, C9SM02306B, marenduzzo07steady}) with the advection-diffusion equation for the activity field. We assume that this field is the product of the concentration (or quantity) of nutrients, which may vary in space and time, and a constant that measures their efficiency (or quality) in the generation of active stresses. Clearly, when the concentration is constant the activity field is also constant. Note that we assumed the active stress is proportional to the concentration of fuel for simplicity but other choices are possible; it can be proportional to the chemical potential of fuel for instance~\cite{J_licher_2018}.
Here we do not consider the consumption of nutrients (as done, for instance, in Ref.~\cite{PhysRevE.102.020601}) as we are interested in their dispersion throughout the system, i.e., we assume that the nutrients are mixed on a time scale that is much shorter than that of their consumption. As the active stresses drive spontaneous flows in the active turbulent domain and the flow disperses the activity into the passive nematic domain, there is a non-trivial dynamical behaviour at this active-passive nematic interface. Note that the interface evolves but its propagation is constrained by the conserved activity. It can not propagate indefinitely with constant velocity as in Ref.~\cite{C9SM02306B} which does not consider the conservation of activity. 

At the coarse grained level, the nematic is described by the director field $n_\alpha$, representing the average direction of molecular aligning, and the scalar order parameter $S$, which represents the degree of nematic order in a specific domain. These two fields are combined in the uniaxial tensor order parameter, $Q_{\alpha \beta} = S(n_\alpha n_\beta - \delta_{\alpha \beta}/3)$, a traceless and symmetric tensor. For simplicity, we consider the tensor $Q_{\alpha\beta}$ uniaxial, which is exact in bulk passive nematics and a good approximation under most conditions. The equilibrium of the system is described by the Landau-de Gennes free energy $\mathcal{F} = \int_V \,d^3 r\, f$, with energy density given by:
\begin{align}
 f=& \frac{A_0}{2}\left( 1- \frac{\gamma}{3} \right) {Q_{\alpha \beta}}^2 - \frac{A_0\gamma}{3} (Q_{\alpha \beta} Q_{\beta \gamma} Q_{\gamma \alpha})   \nonumber\\
 &+ \frac{A_0\gamma}{4} {Q_{\alpha \beta}}  ^4 + \frac{K}{2} (\partial _\gamma Q_{\alpha \beta})^2 .
\end{align}
Here $A_0$ is a positive constant that sets the scale of the bulk free energy, $K$ is the single elastic constant, where we neglected elastic anisotropy (see Ref.~\cite{doi:10.1098/rsta.2020.0394} for the effect of elastic anisotropy), and $\gamma$ is a function of the field that drives the ordering transition, such as temperature for thermotropic nematics. At coexistence between the isotropic ($S=0$) and the nematic ($S=S_N$) phases, $\gamma$ is $2.7$ and the nematic order parameter is  $S_N=1/3$. The simulations reported below are in the deep nematic phase, with $\gamma=3$ and $S_N=1/2$.

The time evolution of the system is governed by the Beris-Edwards~\cite{beris1994thermodynamics}, the continuity, the Navier-Stokes~\cite{beris1994thermodynamics} and the advection-diffusion equation~\cite{succi2018lattice}, respectively: 
\begin{align}
  &\partial _t Q_{\alpha \beta} + u _\gamma \partial _\gamma Q_{\alpha \beta} - S_{\alpha \beta} = \Gamma H_{\alpha\beta} , \label{beris-edwards-eq} \\
&  \partial _\beta u_\beta = 0, \label{continuity-eq}\\
&\rho\partial_t  u_\alpha + \rho u_\beta \partial _\beta   u_\alpha  = -\chi u_\alpha + \partial_\beta [ 2\eta D_{\alpha\beta}  + \sigma^{\text{lc}}_{\alpha\beta} -\zeta C Q_{\alpha\beta} ],   \label{navier-stokes-eq}\\
&\partial _t C + \partial _\beta (  C u_\beta ) = D_m \partial_\beta  \partial _\beta C, \label{adv-diff-eq}
\end{align}
where $D_{\alpha\beta} = (\partial_\alpha u_\beta + \partial_\beta u_\alpha)/2$ is the shear rate.
Equation.~\eqref{beris-edwards-eq} describes the evolution of the order parameter $Q_{\alpha\beta}$ for  nematics,  Eqs.~\eqref{navier-stokes-eq} and~\eqref{continuity-eq} describe the evolution of the velocity field $u_\alpha$ and Eq.~\eqref{adv-diff-eq} describes the evolution of the activity field. $\Gamma$ is the system dependent rotational diffusivity, $\rho$ is the density, $\chi$ is the friction with the substrate, $\eta$ is the shear-viscosity and $D_m$ is the diffusion coefficient. Although the Reynolds number is typically small in active turbulent systems, the inertial term in Eq.~\eqref{navier-stokes-eq} can influence the flow as its effect accumulates in time and may give rise to large-scale fluid motion~\cite{PhysRevLett.127.268005}. Moreover, while we use a 3D model the directors lie within the plane and the gradients in the third direction, perpendicular to the plane, are zero as we apply periodic conditions in this direction. Under these conditions, the model becomes effectively 2D~\cite{C9SM02475A}. The last term in Eq.~\eqref{navier-stokes-eq} is the active stress, which corresponds to a force dipole density, with the activity strength $\zeta$ being positive for extensile systems (or pusher particles) and negative for contractile ones (puller particles). $C$ is the local concentration of nutrients modelled by a scalar field that is globally conserved. We refer to the product $\zeta C$ as the activity field in line with the nomenclature used in previous works~\cite{C9SM00859D, C9SM02306B, marenduzzo07steady}. Thus, $\zeta$ is the coupling constant between the concentration of nutrients and the generation of active stresses (it can be thought of as the ``quality'' of the nutrients or the effect of a particular nutrient on different species of bacteria) while $C$ is the concentration of those nutrients. Gradients in $Q_{\alpha\beta}$ and $C$ produce a flow field, which is the source of the hydrodynamic instabilities in active nematics. The co-rotational term is: 
\begin{align*}
 &S_{\alpha \beta} = ( \xi D_{\alpha \gamma} + W_{\alpha \gamma})\left(Q_{\beta\gamma} + \frac{\delta_{\beta\gamma}}{3} \right) \nonumber \\
&
 + \left( Q_{\alpha\gamma}+\frac{\delta_{\alpha\gamma}}{3} \right)(\xi D_{\gamma\beta}-W_{\gamma\beta}) 
 \nonumber \\
&
 - 2\xi\left( Q_{\alpha\beta}+\frac{\delta_{\alpha\beta}}{3}  \right)(Q_{\gamma\epsilon} \partial _\gamma u_\epsilon), 
\end{align*}
where $W_{\alpha\beta}= (\partial _\beta u_\alpha - \partial _\alpha u_\beta )/2$ is the vorticity. $\xi$ is the flow aligning parameter, which may be related to the shape of the particles of the nematic, being positive for rod-like and negative for disk-like particles.  
The molecular field $H_{\alpha\beta}$ describes the relaxation of the order parameter towards equilibrium:  
\begin{align}
 H_{\alpha\beta} = -\frac{\delta \mathcal{F}}{\delta Q_{\alpha\beta}} + \frac{\delta_{\alpha\beta}}{3} \Tr \left( \frac{\delta \mathcal{F}}{\delta Q_{\gamma \epsilon}} \right).
\end{align}
The passive nematic stress tensor is given by the Landau-de Gennes theory~\cite{beris1994thermodynamics}:
\begin{align} 
 \sigma_{\alpha\beta}^{\text{lc}} &= -P_0 \delta_{\alpha\beta} + 2\xi \left( Q_{\alpha\beta} +\frac{\delta_{\alpha\beta}}{3} \right)Q_{\gamma\epsilon}H_{\gamma\epsilon}  \nonumber \\ 
 & - \xi H_{\alpha\gamma} \left( Q_{\gamma\beta}+\frac{\delta_{\gamma\beta}}{3} \right) - \xi \left( Q_{\alpha\gamma} +\frac{\delta_{\alpha\gamma}}{3} \right) H_{\gamma \beta}   \nonumber \\ 
 & - \frac{\delta \mathcal{F}}{\delta (\partial_\beta Q_{\gamma\nu})}\, \partial _\alpha Q_{\gamma\nu}  + Q_{\alpha\gamma}H_{\gamma\beta} - H_{\alpha\gamma}Q_{\gamma\beta} ,
\end{align}
where $P_0$ is the isotropic pressure.
This system of differential equations is solved using a hybrid method with the same spatial discretization: Eq.~\eqref{beris-edwards-eq} is solved using finite-differences, Eqs.~\eqref{navier-stokes-eq} and~\eqref{continuity-eq} are recovered in the macroscopic limit with the lattice Boltzmann method and Eq.~\eqref{adv-diff-eq} is also solved with lattice Boltzmann for advection-diffusion~\cite{kruger2016lattice}. Notice that Eq.~\eqref{beris-edwards-eq} and~\eqref{navier-stokes-eq} are coupled through $Q_{\alpha\beta}$ and $\mathbf u$ and Eq.~\eqref{navier-stokes-eq} and Eq.~\eqref{adv-diff-eq} through $\mathbf u$ and $C$. 
The results are given in lattice units, where the distance between nodes $\Delta x$, the time step $\Delta t$ and the density $\rho$ are equal to 1. 
Except where otherwise stated we set: kinematic viscosity $\nu = 0.133$, $K=0.01$, $A_0=0.1$, $\Gamma = 0.34$, $\xi=0.7$ (flow aligning regime) and $\chi = 0.1$. We choose a small diffusion coefficient $D_m=3.3\times 10^{-4}$, which corresponds to high Péclet numbers, since we are interested in the regime where advection dominates diffusion. Initially, the maximum concentration $C$ is one and the minimum zero. The boundary conditions are periodic in both directions of the simulation box. For this set of parameters, the nematic correlation length is $\ell_N = \sqrt{27K/(A_0 \gamma)} = 0.95$, which is close to the lattice spacing. The screening length, defined as the ratio between the shear stress and that due to substrate friction, is $\ell_F = \sqrt{\nu/\chi} = 1.15$, is also close to the lattice spacing. These two lengths were kept fixed. The friction force in Eq.~\ref{navier-stokes-eq} is required to model the friction with the substrate as in swarming bacteria or the oil viscosity as in microtubule-kinesin mixtures~\cite{PhysRevX.11.031065}. It introduces one extra length scale, $\ell_F$, and results in smaller vortex areas at high friction coefficients. As discussed in Ref.~\cite{alert082321-035957}, the friction also changes the scaling exponents in the energy spectra.
The active length, $\ell_A = \sqrt{K/\zeta}$, defined as the ratio between the elastic and the active stresses, was varied between $\ell_A = 0.63$ for $\zeta=0.025$ and $\ell_A=1.83$ for $\zeta=0.003$. The active length $\ell_A$ sets the average size of the vortices in the active turbulent regime, which is $\sim 11 \ell_A$ as reported in Ref.~\cite{C9SM02306B}. This choice of parameters was found to be a reasonable compromise between the spatial and temporal resolution of the simulations and long simulation times, allowing us to average over random initial conditions and to explore the effects of geometry, activity and the aligning parameter, as described below. Except for Eq.~\ref{adv-diff-eq} and the value of $\gamma$, the same model and set of parameters was used in Ref.~\cite{C9SM02306B}, where it is shown that this hydrodynamic model captures the main features of swarming bacteria and their active-passive interfaces.

\begin{figure*}[th]
\center
\includegraphics[width=0.9\linewidth]{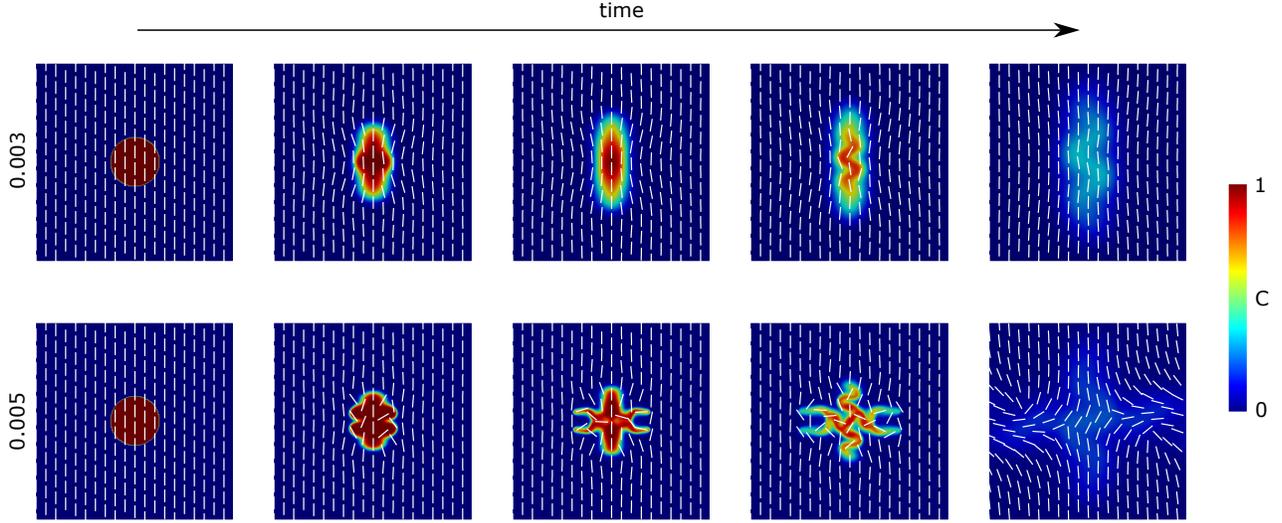}
\caption{Time evolution of an active circular region with a uniform initial concentration of the scalar field $C=1$ inside (red) and $C=0$ outside (blue). The namatic is uniformly aligned everywhere. Domains with different activity strengths are illustrated: $\zeta=0.003$ and $\zeta=0.005$. The white bars indicate the directors. The time interval between the frames is not constant.}
\label{circle-fig}
\end{figure*}
\begin{figure*}[th]
\center
\includegraphics[width=0.8\linewidth]{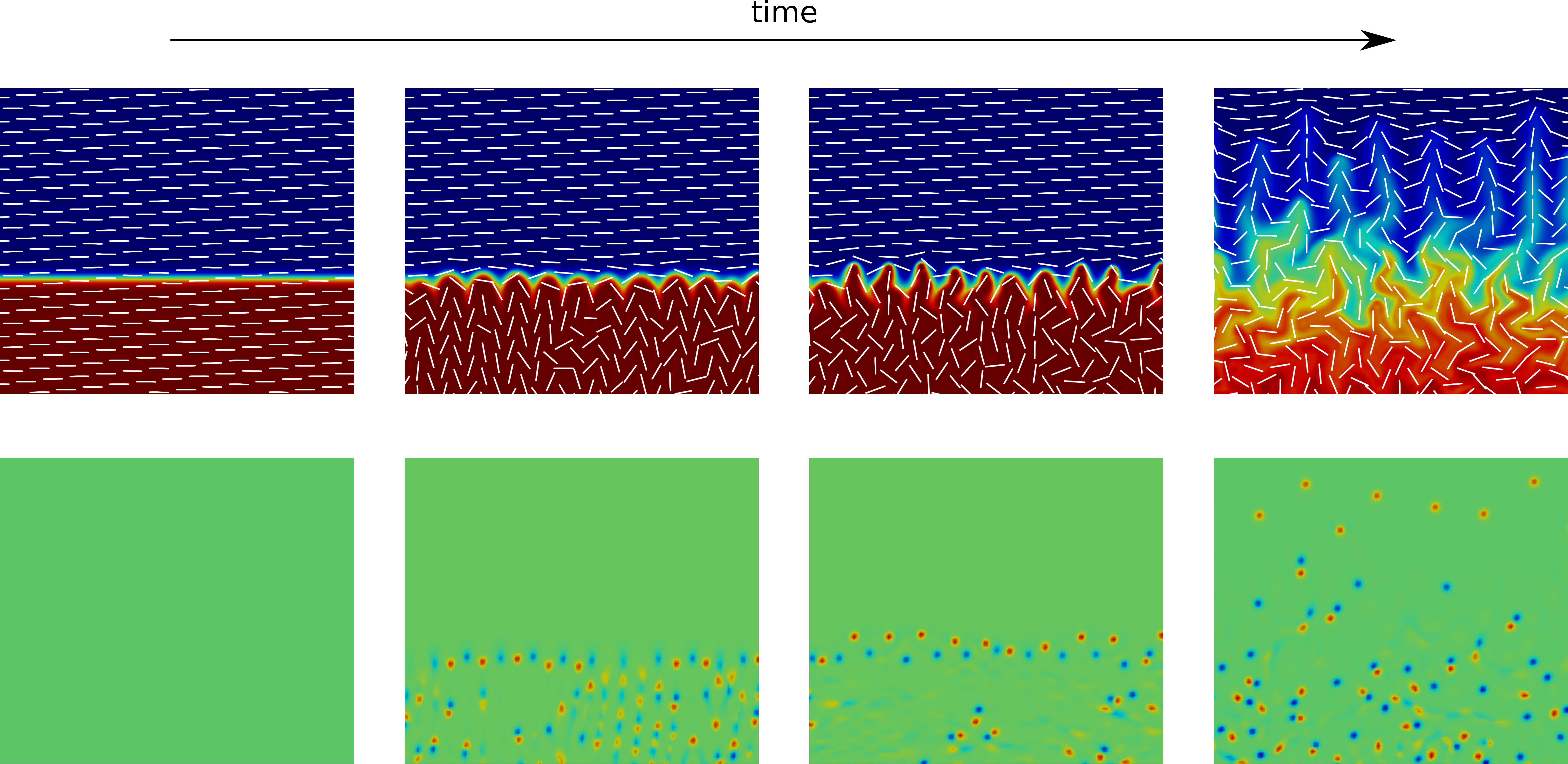}
\caption{Time evolution of an initial step-like activity profile for a system with $\zeta = 0.009$. A small region of the $L_X\times L_Y=800 \times 400$ domain is shown. On the top row, the colors represent the concentration profiles with red being the initial concentration $C=1$ and blue being $C=0$. The white bars indicate the directors. On the bottom, the coloured dots represent the charge of the defects, with red being positive (comet) and blue being negative (trefoil) defects. The time interval between the frames is not constant.}
\label{flat-fig}
\end{figure*}

\section{Sharp active-passive interfaces}
\label{sharp-sec}
In this section, we study the early stages of the evolution of sharp active-passive interfaces in two simple geometries: circular and  flat. Both the active and passive domains are initially ordered.

In previous works~\cite{C9SM00859D, doi:10.1098/rsta.2020.0394}, we considered the active force at the nematic-isotropic interface of systems with homogeneous activity. Here we consider this force at the active-passive nematic interface where the scalar order parameter $S$ is roughly uniform, i.e. both phases are locally ordered, with a heterogeneous activity field. The interface is characterized by an evolving activity profile that decreases from high to low values across the interface. During the evolution, the initially sharp interface will exhibit bend (splay) instabilities for extensile (contractile) systems and localized defects where the order parameter $S$ is zero~\cite{PhysRevLett.113.248303, C7SM00325K}. The active force is the divergence of the active stress: $F^a_\alpha =  -\partial_\beta [\zeta C(\mathbf{x}) Q_{\alpha \beta}(\mathbf{x})]$. Using the definition of $Q_{\alpha\beta}$, we find:
\begin{align}
F_\alpha^a = - S \zeta \Big [    n_\alpha n_\beta \partial_\beta C - \frac{\partial_\alpha C}{3} + C n_\alpha \partial_\beta n_\beta + C n_\beta \partial_\beta n_\alpha    \Big ].
\label{active-force-eq}
\end{align}
Let us focus on the component of this force perpendicular to the active-passive interface, where the normal vector is $\mathbf{m} \equiv -\nabla C / \vert  \nabla C \vert$. This interface is easily located by the abrupt change in the activity field ($\zeta C$), which is clear in the initial stages of the simulations reported in this section. Thus, the perpendicular component of the force is:
\begin{align}
\mathbf{F}_\perp^a &= - S \zeta \Big[  (\mathbf{n} \cdot \mathbf{m}) (  \mathbf{n} \cdot \nabla C + C \nabla \cdot \mathbf{n} ) \nonumber \\
 & + C\mathbf{m}\cdot (\mathbf{n}\cdot \nabla) \mathbf{n}  - \frac{1}{3}\mathbf{m}\cdot \nabla C   \Big] \mathbf{m}.
\label{perp-force}
\end{align}
Notice that this expression is similar to that obtained in Ref.~\cite{C9SM00859D} for a nematic-isotropic interface, with $S$ replaced by $\zeta C$.

\subsection{Active circle}

We consider a nematic with a uniform director field pointing in the vertical direction and a circular region of radius $R=38$ with uniform concentration $C=1$ and positive activity (extensile system) surrounded by a passive nematic ($C=0$). The velocity is initially zero, and the dimensions of the simulation box are $L_X \times L_Y = 200 \times 200$. Using Eq.~\eqref{perp-force}, we find for the perpendicular force on the top and bottom and on the right and left, respectively:
\begin{align}
\mathbf{F}^{\rm tb}_\perp = \frac{2S \zeta \vert \nabla C \vert}{3} \mathbf{m}, \quad \mathbf{F}^{\rm rl}_\perp =     - \frac{S \zeta \vert \nabla C \vert}{3} \mathbf{m}.
\end{align}
Thus, the active force points outwards on the top and bottom and inwards on the right and left, elongating the circle vertically for positive activities. Note that this elongation as well and the other instabilities that are observed are due to the coupling between the concentration $C$ and the director$\mathbf{n}$ fields. The active force acts on the velocity field (see Eq.~\ref{navier-stokes-eq}) and it becomes zero when the gradients in $C$ and $\mathbf{n}$ are negligible. 
Figure~\ref{circle-fig} illustrates the time evolution of this active circle for different activity strengths in the circular region. As the concentration $C$ profile evolves due to advection-diffusion, we will either refer to the initial concentration for the homogeneous case or to the largest initial concentration for the heterogeneous one, which are equal to one. The largest initial activity corresponds to the maximum of the activity field, which is $\zeta$.
We find that, for $\zeta=0.003$, the circle elongates vertically as expected. After some time, the opposite forces on both sides destabilize the straight active region through an interfacial bend instability similar to that observed in multicomponent systems~\cite{doi:10.1098/rsta.2020.0394, PhysRevLett.113.248303}. For larger activities, $\zeta = 0.005$, the circle elongates initially, and then a new instability sets in the form of two jets from either side of the active domain, which propagate in almost straight trajectories. These jets are formed by motile positive (or comet-like) defects as will be discussed later. In both cases, the scalar activity field is dispersed and diluted over larger areas of the domain. As we will see in Sec.~\ref{dispersion-sec}, the dispersion is faster at higher activities due to the larger velocities of the comet-like defects. We note that this dispersion is not isotropic by contrast to diffusion. While these jets promote the dispersion of activity, there is also an active force compressing the active region, which hinders its dispersion. Similar jets have been observed in experiments with bacterial baths~\cite{PhysRevLett.84.3017}. 

Eventually, the activity disperses throughout the entire domain and becomes homogeneous. If the final concentration is small, the system becomes distorted but static where the flows are screened by the friction force~\cite{Doostmohammadi2016}. At higher final activities, the whole system becomes active turbulent with a constant rate of creation and annihilation of pairs of defects. Although these final states may be interesting, we did not investigate them further since the assumption that the activity field is conserved is unlikely to be realistic at very long times, as its consumption will eventually set in.

\subsection{Flat interface}
\label{flat-sec}

We proceed to consider a flat active-passive interface with uniformly aligned directors parallel to the interface. We perturb the initial directors with random uniformly distributed perturbations up to $1^\circ$ and use a domain with dimensions $L_X \times L_Y = 800 \times 400$ in the simulations that follow. The concentration is initially $C=1$ at the center ($L_X/2-75 < y < L_X/2+75$) and $C=0$ elsewhere. There are two interfaces as a result of the periodic boundary conditions. We start by analysing the active force at the interface. From Eq.~\eqref{perp-force}, we find that the perpendicular force becomes
 \begin{align}
 \mathbf{F}_\perp^a = - S \zeta \Big[   C\mathbf{m}\cdot (\mathbf{n}\cdot \nabla) \mathbf{n}  - \frac{1}{3}\mathbf{m}\cdot \nabla C   \Big] \mathbf{m}.
 \end{align} 
The term proportional to $\nabla C$ does not change when the interface undulates, but the first term $(\mathbf{n}\cdot \nabla) \mathbf{n}$ changes sign with that of the interfacial curvature. This results in a larger interfacial force inwards than outwards in the active region. Again, this is analogous to what happens in multicomponent systems~\cite{PhysRevLett.113.248303}. In Fig.~\ref{flat-fig}, on the top, we depict the time evolution of the flat interface in a small region of the domain at the top interface. In the initial stages, the interface undulates almost periodically. As the inward active force increases, the curvature of the concave interfacial regions becomes higher than that of the convex ones. Then the convex regions become jets (motile positive defects) which propagate in the direction perpendicular to the interface while the concave regions become quasi-stationary negative defects.

\begin{figure}[th]
\center
\includegraphics[width=\linewidth]{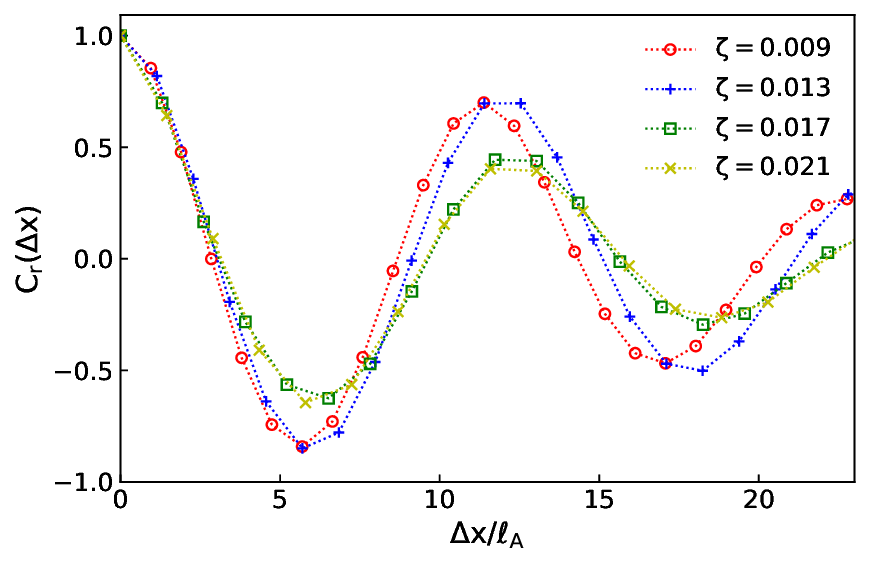}
\caption{Spatial correlation of the active-passive interfacial height in the early stages (when the standard deviation of the interfacial height is less than $\sigma=1$) for different activity strengths. The $x$-axis was scaled by the active length $\ell_A$. }
\label{correlation-fig}
\end{figure}
\begin{figure}[th]
\center
\includegraphics[width=\linewidth]{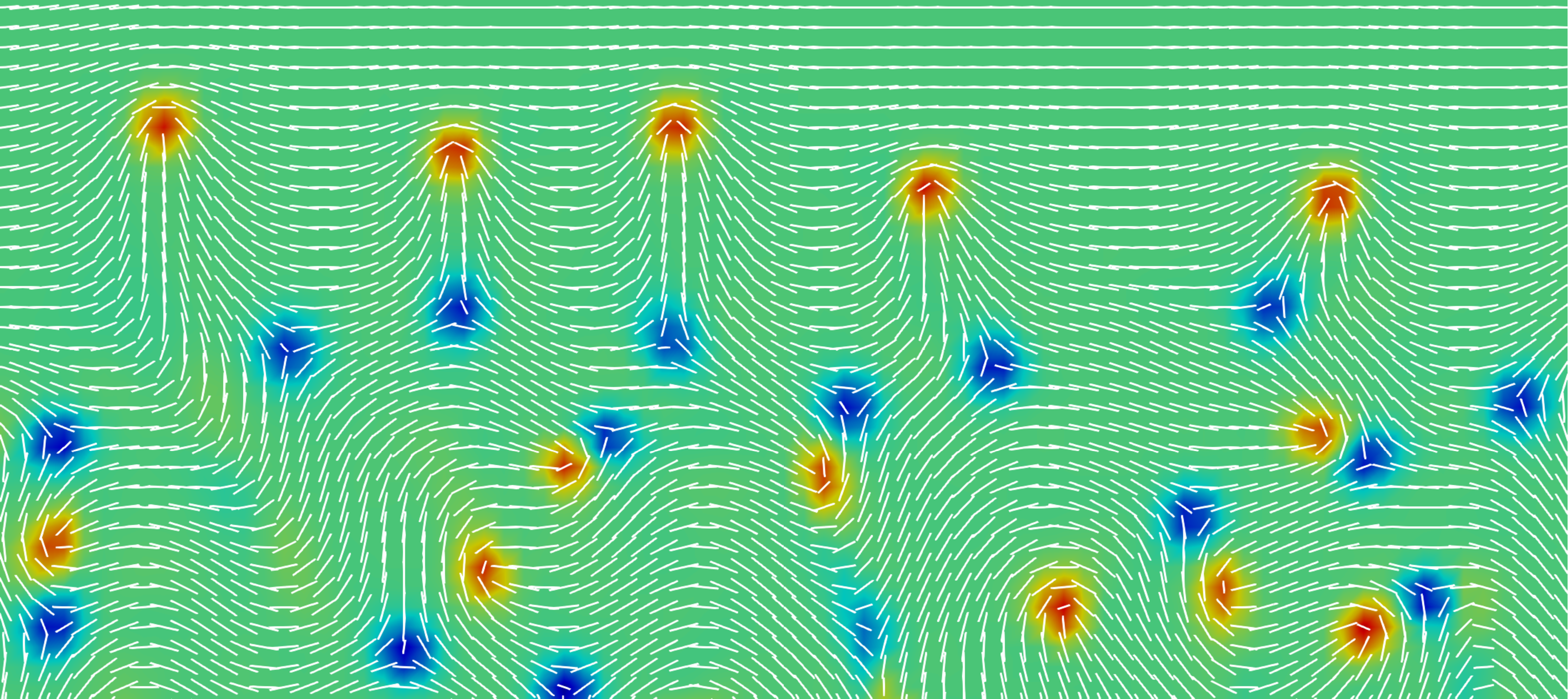}
\caption{Charge density of a region close to the flat interface corresponding to Fig.~\ref{flat-fig} showing the director field configuration close to the defects. The colored dots represent the charged  defects, with red being positive (comet) and blue being negative (trefoil) defects. The white bars illustrate the directors (one per lattice node).}
\label{charge-flat-fig}
\end{figure}
\begin{figure}[th]
\center
\includegraphics[width=\linewidth]{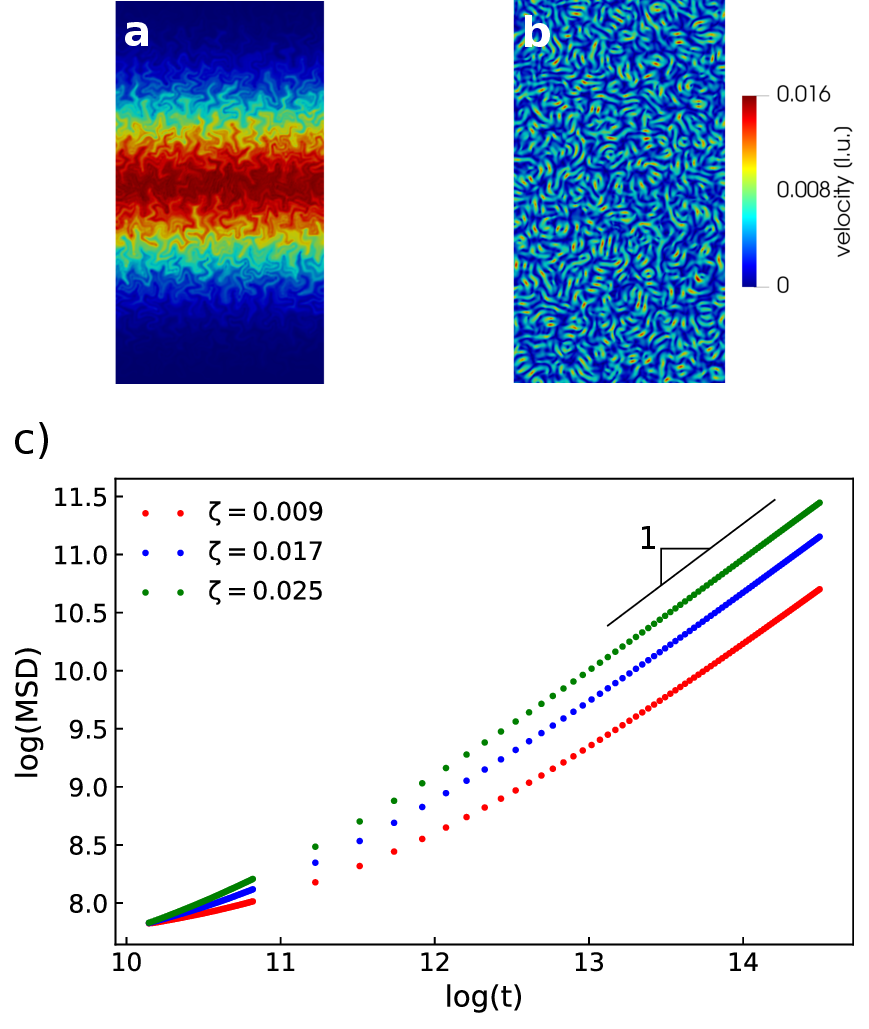}
\caption{Dispersion of an inert solute (unable to produce active stresses) by the active turbulent nematic with different uniform activities. a) Concentration field, where red represents the largest initial concentration $C_s=1$ and blue stands for $C_s=0$. The width of the region shown in the figure is $L_X$ while the height was reduced as it is mostly isotropic. b) Velocity field in the active nematic at the time and in the region corresponding to ``a''. c) Mean square displacement of the concentration field $C_s$ over time. The solid line has a slope equal to one.}
\label{passive-fig}
\end{figure}

The wavelength of the initial interfacial undulations decreases as the activity increases due to the interplay between the elastic and the active forces. This effect is quantified by the active length, $\ell_A$.
We calculate the spatial correlation function of the interfacial height $h(x)$ for different activities:
\begin{equation}
C_r(\Delta x) = \left \langle  \frac{h(x_0)h(x_0+\Delta x)}{ h(x_0)^2} \right \rangle,
\end{equation}
where the average is over the reference position $x_0$ and time.
This was done at the early stages when the undulation instability sets in, defined as the time when the standard deviation of the interfacial height is $\sigma_y=1$. The interfacial position is defined by the nodes where the concentration is $C=0.5$, i.e., one half of its initial value. Fig.~\ref{correlation-fig} shows that the first minimum of the correlation function collapses if the distance is scaled by the active length. The characteristic wavelength of the interfacial undulations may be estimated from the position of the second zero of the correlation function, which is $\sim 10 \ell_A$. This length is close to the average size of the vortices in the active turbulent regime calculated in Ref.~\cite{C9SM02306B} ($\sim 11 \ell_A$). 
The size of both structures is determined by the balance between the active and elastic forces but the interfacial instability occurs at lower activities than the active turbulent regime, since the latter is screened by the finite domain size and the friction with the substrate when the interfacial instability occurs. Notice that, although the active length is of the order of the lattice spacing, the director and velocity fields change over a length scale that is one order of magnitude larger.

In order to analyse the dynamics of the defects, we calculate the charge density as~\cite{PhysRevLett.113.248303}:
\begin{align}
 q = \frac{1}{4\pi} \left( \partial _x Q_{x\alpha} \partial_y Q_{y\alpha} - \partial _x Q_{y\alpha}\partial_y Q_{x\alpha}  \right).
\end{align}
On the bottom row of Fig.~\ref{flat-fig}, the charge density field is plotted for the regions and times depicted on the top row. Initially, the charge density is zero everywhere and, as the interfacial instability develops, pairs of positive ($+1/2$) and negative ($-1/2$) defects are created along the interface and in the bulk. Fig.~\ref{charge-flat-fig} illustrates the director field configuration near the defects.
The positive defects created at the interface move towards the passive region while the negative ones remain quasi-stationary in their original positions. There is a clear segregation of positive and negative defects at the active-passive interface in line with the results reported in Ref.~\cite{PhysRevX.9.041047} for a static heterogeneous activity profile. Reference~\cite{doi:10.1073/pnas.2106038118} reports the experimental observation of defects segregation at high-low friction interfaces, which were designed by changing the depth of micropatterned structures. The main difference between these static interfaces and the dispersion of activity discussed here is that the segregation of positive and negative defects reach a steady state in the former while defects keeps separating in the latter until the activity becomes homegeneous.
The positive defects that leave the active region form jets which propagate in a direction perpendicular to the interface until the activity carried by them is dispersed. The direction of propagation of the defects is in line with previous theoretical predictions, which considered active torques on individual defects~\cite{Zhang2021, PhysRevX.9.041047}. The negative defects left behind, in the interfacial region, form a barrier which hinders newly formed jets. Most of the newly created positive defects that try to leave the active region are annihilated by one of the negative defects at this barrier (this annihilation is illustrated in Fig.~\ref{defects-fig} for a Gaussian activity field). Thus, the interfacial barrier of negative defects hinders the dispersion of activity in two ways: first, they do not move and, therefore, do not disperse activity; second, they annihilate newly created positive defects which are responsible for most of the dispersion. In the initially active region, the flow develops active turbulence after a few thousand time steps.

In the active circle of Fig.~\ref{circle-fig}, the motile positive defects are formed on the sides of the circle at high activities,  $\zeta=0.005$, while the negative defects remain in the center. Note that the undulations that lead to the emergence of the jets in the active circle are similar to the undulations that occur at the flat interface, with a wavelength $\sim 10 \ell_A$. In the active circle with $\zeta=0.003$, this wavelength corresponds to $18.3$, which is of the order of the radius of the initial circle and thus no jets are observed. At higher activities (smaller active lengths) the wavelength of the undulations becomes smaller and jets are formed, as at the flat interface.

\section{Dispersion of the activity}
\label{dispersion-sec}

In this section, we investigate how the dispersion of the activity field depends on the activity strength and on the aligning parameter of the system. For comparison, we start by calculating the dispersion of an inert solute which is passively transportated (i.e. it does not produce active stresses) by an active turbulent fluid (which is rendered active by a distinct source of activity). 

\begin{figure}[th]
\center
\includegraphics[width=\linewidth]{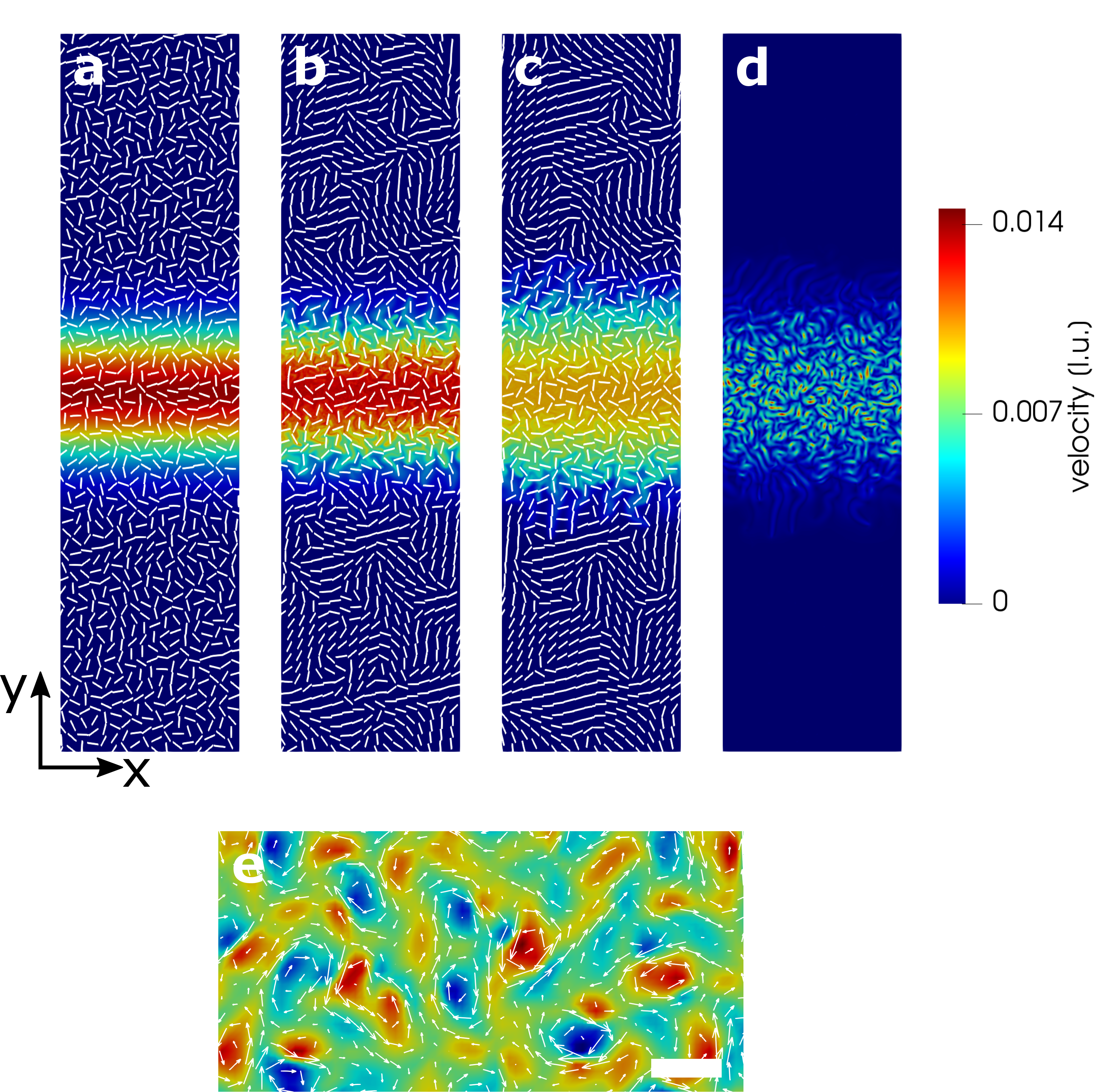}
\caption{Dispersion of a Gaussian activity field. Figures a), b) and c) show the concentration (colors) and director (lines) fields at three different times in lattice units: $0$, $36000$ and $52000$ respectively. d) Velocity field corresponding to figure ``c''. e) Small region  in the center of figure ``d'' with the z-component of the vorticity (colors) and velocity field (arrows). The scale bar is $10 \Delta x$.}
\label{screenshots-fig}
\end{figure}
\begin{figure}[bh]
\center
\includegraphics[width=\linewidth]{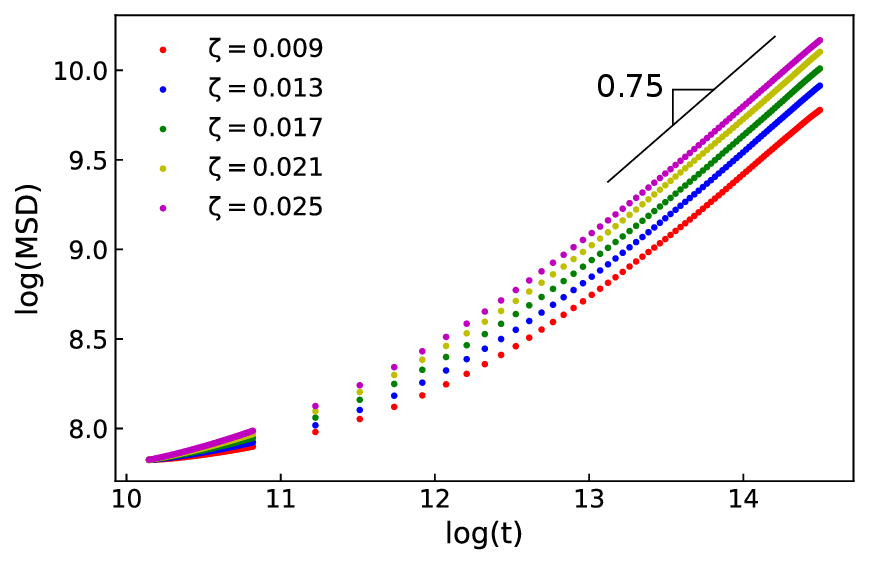}
\caption{Mean square displacement of the scalar field $C$ over time for different activity strengths. The solid line has a slope $0.75$. }
\label{msd-zeta-fig}
\end{figure}

\subsection{Dispersion of an inert solute by an active turbulent nematic with uniform activity}
\label{passive-sec}

We simulated the dispersion of an inert solute (unable to produce active stresses) in an active turbulent nematic with uniform activity. This is an intermediate step to be used as a reference for the dispersion of the activity field reported later on. We simulate a homogeneous and static activity in a domain of dimensions $L_X \times L_Y = 200 \times 800$ with periodic boundary conditions. These boundary conditions, used in all the simulations reported here, guarantee the conservation of the dispersed quantity. To simulate the dynamics of this inert solute, we consider that the concentration of nutrients is $C=1$ (constant and homogeneous) in Eq.~\eqref{navier-stokes-eq}. The concentration of the inert solute, called $C_s$, is allowed to evolve according to Eq.~\eqref{adv-diff-eq}. As a result these equations are decoupled in this section and the source of activity is not connected to the inert solute concentration $C_s$.
The initial conditions, correspond to directors set at random directions and fluid at rest. We then let the system evolve to an active turbulent state for the first $25000$ iterations. For the advection-diffusion equation, we initialize the concentration field of the inert solute $C_s$ as a Gaussian along $y$ centred at $y=L_Y/2$ with standard deviation $\sigma=50$. This Gaussian field is allowed to evolve only after $t=25000$ steps, when the nematic is in the turbulent regime.  Figure~\ref{passive-fig} (a) and (b) illustrate the concentration of the inert solute and the velocity fields at a given time ($t=37500$) in a small central region of the domain. The velocity field exhibits a number of vortices in the entire domain since the activity is constant, as in Ref.\cite{C9SM02306B}. In active turbulent systems, the characteristic velocity is expected to increase with the activity as $v_{ch} \sim \sqrt{\zeta}$~\cite{C6SM00812G}. Thus, we also expect the dispersion of the inert solute field (powered by the advection) to increase with the activity. We measured the mean square displacement (MSD) of the inert solute concentration field from its second moment:
\begin{align}
 {\rm MSD} = \frac{1}{C_{\rm s, tot} } \sum_{i} C_s(y_i) (y_i-\langle y\rangle )^2 ,
  \label{msd-eq}
\end{align}
where
\begin{align}
 \langle y\rangle = \frac{1}{C_{\rm s, tot}} \sum_i C_s(y_i) y_i,
\end{align}
$C_{\rm s,tot}$ is the total amount of the inert solute $C_s$ and the sum runs over all the $i$ nodes of the square lattice. In Fig.~\ref{passive-fig}(c), we plot the logarithm of the MSD versus the logarithm of time for three different activities. After an initial transient, the dispersion evolves to a diffusive regime, characterized by a slope equal to 1 as shown in the figure. This implies that the motion of an element of fluid containing the concentration of the inert solute does not affect the fluid velocity, which is random in the active turbulent nematic with constant activity. The MSD increases with the activity as expected.

\begin{figure*}[ht]
\center
\includegraphics[width=\linewidth]{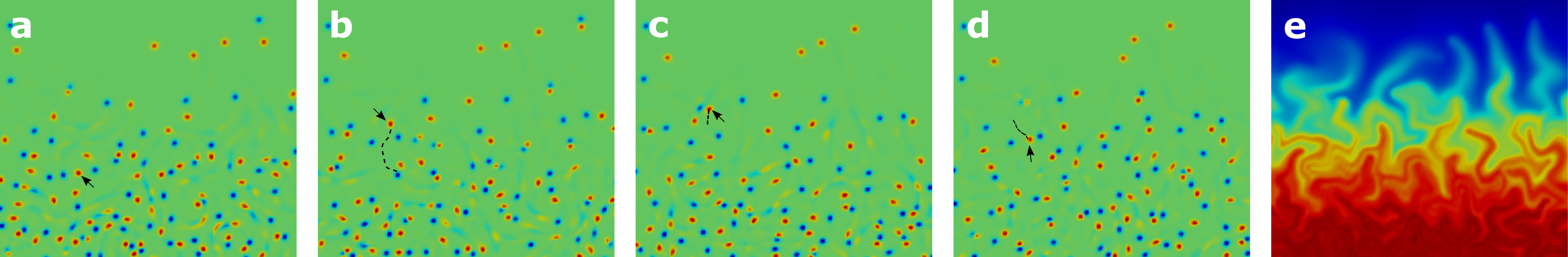}
\caption{Motion of the defects for a Gaussian activity field with maximum initial concentration $C=1$ and activity strength $\zeta=0.009$. Figures a) to d) show the time evolution of the defects in a small region of the $200 \times 800$ domain. Red indicates the positive defects while blue indicates the negative ones. a) The arrow marks the position of one particular defect the trajectory of which (represented by dashed lines) is followed up to $t=36000$. b) The positive defect indicated in ``a'' passes through the barrier of negative defects. c) At $t=41500$, a new pair of defects is created (within the white circle) close to the trajectory of the marked positive defect. d) At $t=43500$, the marked defect is annihilated by the recently created negative defect and the remaining positive defect moves towards the active region. e) Concentration field in the region and time corresponding to figure ``d''.}
\label{defects-fig}
\end{figure*}
\begin{figure}[ht]
\center
\includegraphics[width=\linewidth]{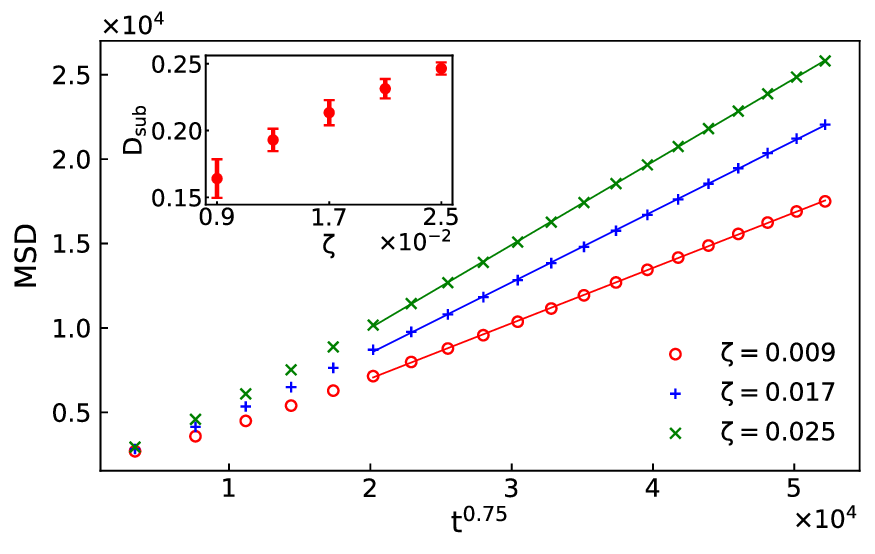}
\caption{Mean square displacement of the scalar concentration of nutrients $C$ versus time$^{0.75}$ giving the generalized diffusion coefficient $D_{sub}$ for different activity strengths (inset). The error bars result from the average over 20 samples.}
\label{ds-zeta-fig}
\end{figure}

\subsection{Activity}
\label{activity-sec}

If the dispersed quantity is the activity itself the dynamics becomes more complex. We performed simulations with initial configurations consisting of a Gaussian concentration field $C$ in the center of the domain, with standard deviation $\sigma =50$ and maximum $C=1$. The initial directors are randomly oriented. The choice of an initial Gaussian concentration field leads to the presence of concentration gradients everywhere in the active region, which maximizes its dispersion. Note that in the configurations discussed in Sec.~\ref{sharp-sec} the dispersion occurs only at the interface. The Gaussian field is held static until $t=25000$ when active turbulence is observed close to the center of the domain. Both active and passive regions are deep in the nematic phase. After $25000$ steps, the Gaussian activity field evolves according to Eq.~\ref{adv-diff-eq}, which disperses the scalar concentration field to  passive nematic regions rendering them active. Figure~\ref{screenshots-fig} (a) to (c) illustrates the time evolution of the scalar field while (d) depicts the velocity field at a given instant of time and (e) the vorticity field of a smaller region in the active domain highlighting the typical vortex size. Although they are less noticeable in a random configuration of directors, there are also jets from the active to the passive regions as in the aligned systems reported in Sec.~\ref{sharp-sec}. We measured the MSD of the concentration field $C$ as in Eq.~\eqref{msd-eq}. In Fig.~\ref{msd-zeta-fig}, we report the MSD of $C$ as a function of time for different activity strengths. Each curve is the average of the MSD of 20 samples with different initial configurations of the random director field. After a transient regime, the MSD evolves with a slope less than one in a log-log plot, for all activities, which reveals the existence of a subdiffusive regime. We contrast this with the results reported in Sec.~\ref{passive-sec} for a system with constant activity, where the dispersion of the concentration of an inert solute was found to be diffusive (slope one). We measured the slope, in the subdiffusive regime, at times between $5\times 10^5$ and $2\times10^6$, for systems with different activities and found an average value $\alpha=0.75 \pm 0.01$, also shown in the figure.

As discussed in Sec.~\ref{flat-sec} for the flat and sharp interface, jets formed by the positive defects move from the active to the passive regions promoting the dispersion of the scalar concentration of activity field. However, the negative defects left behind form a barrier which hinders its dispersion by suppressing the motion of new jets, through annihilation of the defects. This mechanism is still at play in systems with initial inhomogeneous scalar fields and it is responsible for the subdiffusive regime.

Beyond the initial Gaussian concentration field there is another difference from the systems considered in Sec.~\ref{flat-sec}, namely the initial configuration of the directors is random and as a result domains with sizes of the order of $50 \Delta x$ are formed in the passive nematic region (see Fig.~\ref{screenshots-fig}). Note that the size of these ordered domains is much larger than the nematic correlation length or indeed any other physical length in the system (except of course the size of the simulation box). Thus, it is not surprising that the domain structure of this passive nematic region has little or no effect on the dynamics of the dispersion of the activity field. In particular, we note that the size of the ordered domains is much larger than the width of the jets ($\sim \ell_N = 0.95$). We find that the jets still move in the direction of the concentration gradient as in Sec.~\ref{flat-sec}, where the initial condition of the directors was uniform. The dynamics of the motile positive defects is affected when they meet and annihilate with a negative defect in the passive nematic region, and this is not affected by the domain size as it is much larger than the other relevant length scales.  

Figure~\ref{defects-fig} illustrates the motion of the defects close to the Gaussian activity field. Most of the positive defects that move from the active to the passive region are annihilated by negative ones. Even some of the defects that manage to pass through the negative defect barrier are captured by newly created pairs of defects close to the jets trajectories, as indicated by the arrows in Fig.~\ref{defects-fig}. Thus, the barrier of negative defects suppresses the dispersion of scalar activity field and leads to the subdiffusive regime. We expect that the subdiffusive regime does not depend on the initial configuration of the director field. Experimentally, it may be easier to consider a system with homogeneous activity where the active turbulent regime is fully developed and then switch off the activity in the desired passive region.

We define and measured the generalized diffusion coefficient as the proportionality constant between the MSD and $t^\alpha$:
\begin{align}
 {\rm MSD} = 2 D_{\rm sub} t^\alpha.
\end{align}
Figure~\ref{ds-zeta-fig} reveals that the MSD is linear as a function of $t^{0.75}$, and thus we find $D_{\rm sub}$ from a simple linear fit. The inset of Fig.~\ref{ds-zeta-fig}, reveals that the dispersion (or $D_{\rm sub}$) increases with the activity strength. This means that in systems with stronger activities, the dispersion is higher, as expected, since the characteristic velocity of the active turbulent regime is also higher.

\subsection{Aligning parameter}

We proceed to analyse the dependence of the generalized diffusion coefficient $D_{\rm sub}$ on the aligning parameter, which is related to the shape of the particles. The shape can influence many different aspects of the dynamics~\cite{doi:10.1146/annurev.fl.24.010192.001525, Saintillan2015}. For instance, elongated active particles accumulate close to a solid wall forming droplets while spherical particles do not~\cite{zhan_sardina_lushi_brandt_2014, https://doi.org/10.48550/arxiv.2206.15270}. Moreover, simulations of growing bacterial colonies revealed that the cells mix more efficiently when they are elongated~\cite{Schwarzendahl2022}. It is known that bacteria can change their aspect ratio depending on the conditions~\cite{doi:10.1128/JB.182.21.5990-5996.2000, Li777}. It is also possible to control the aspect ratio of bacteria in experiments, which revealed that the vorticity in swarming bacteria is maximized for an intermediate aspect ratio~\cite{Beer2019}. Since the dispersion increases with vorticity, this raises the question of weather bacteria may change their shape in order to increase the dispersion of nutrients throughout the colony. 

\begin{figure}[ht]
\center
\includegraphics[width=\linewidth]{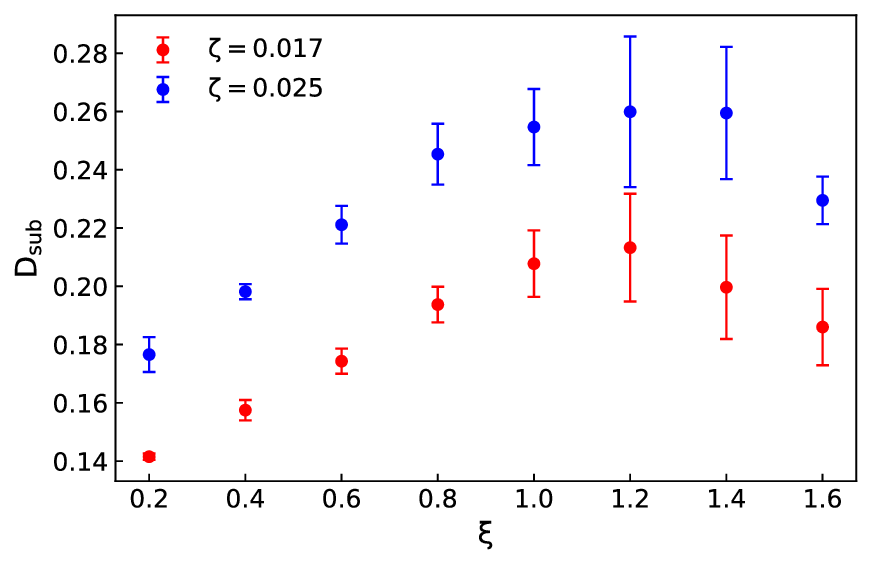}
\caption{Generalized diffusion coefficient as a function of the flow aligning parameter for two different activity strengths (Gaussian profile).}
\label{ds-xi-fig}
\end{figure}

We simulate the system considered in Sec.~\ref{activity-sec} varying the aligning parameter $\xi$ for the same activity strength. In Fig.~\ref{ds-xi-fig}, we plot the generalized diffusion coefficient as a function of the aligning parameter for two different activity strengths. We took averages over 10 samples with different initial random director configurations, which provides the error bars due to the statistical averages. The results reveal that $D_{\rm sub}$ is non-monotonic suggesting the existence of an optimal aspect ratio (related to $\xi$) which maximizes the dispersion of the scalar field $C$. This occurs around $\xi_{max} \approx 1.2$ for both activities. 
As a comparison, passive nematics exhibit a transition at $\xi^\ast =3S_N/(S_N + 2) = 0.6$, for our choice of parameters, with flow tumbling occurring when $\xi<\xi^\ast$ and flow aligning when $\xi>\xi^\ast$. Thus, $\xi_{max}$ is about two times  $\xi^\ast$. In our simulations, we observe significant differences in the jets for different values of $\xi$ although they all move toward the passive region. At small values of $\xi$, the jets change directions more often and thus, move in more diffuse trajectories. At large values of $\xi$, the jets’ trajectories are almost ballistic, but often in oblique directions thus dispersing the activity in directions that are not normal to the interface. In between, at the optimal value of $\xi$, the jets change direction during the trajectory, but move on average perpendicular to the interface. This, in turn, is related to how easily the directors change due to the gradients in velocity.  Although, as far as we know, there are no models that relate the aligning parameter to the aspect ratio of the particles (models exist for the flow tumbling~\cite{larson1999}, but not for the flow aligning regime considered here) the result is interesting by itself. Its biological relevance remains to be established.

\subsection{Dispersion of tracers}

\begin{figure}[h]
\center
\includegraphics[width=\linewidth]{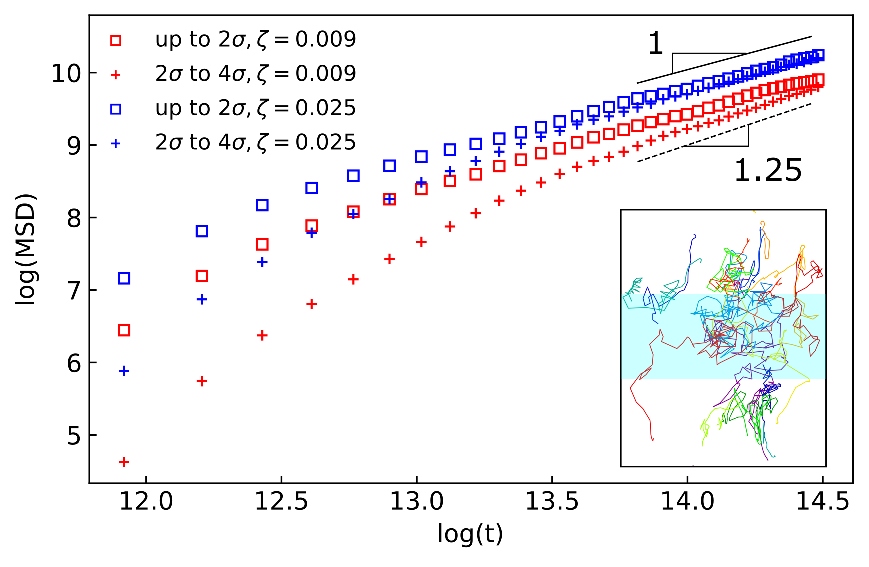}
\caption{Mean square displacement of the position of the tracers versus time in two different initial regions: inner region, where most of the initial activity is concentrated (from the center to $2\sigma$) and outer region (from $2\sigma$ to $4\sigma$ on both sides). The results are shown for two different activity strengths. In both cases, $\sigma=50$ is the standard deviation of the initial Gaussian field $C$. The inset illustrate typical trajectories, which are for a few tracers (30 of 1000) with $\zeta=0.009$. The blue region indicates the inner region (within $2\sigma$) and the lines with different colors represent the trajectories of different tracers. }
\label{tracer-fig}
\end{figure}

Finally, we measured the MSD of tracers randomly placed in the domain. The tracers represent the trajectories of individual particles (or bacteria) in the flow and can be analysed in different initial regions of the domain. They also provide a more accessible way to observe signatures of the heterogeneity of the activity field in experiments. In the systems considered in Sec.~\ref{activity-sec}, with a Gaussian initial profile of the activity field, we placed 1000 tracers randomly throughout the domain, in 20 independent samples, and evolved their positions by assuming that they are massless and are simply advected by the flow, i.e., they have the same velocity as the fluid. It is also assumed that the tracers do not interact as they are scattered and will rarely collide. In this sense, their spreading is fundamentally different from that of the passive solute in Sec.~\ref{passive-sec} as the solute will also spread by diffusion if there are gradients in the concentration field. We calculated the MSD of the tracers placed in two different regions: a central (inner) region, which includes most of the initial activity (within $2\sigma$) and an outer region (between $2\sigma$ and $4\sigma$ on both sides). The initial activity in the outer region is much smaller (nearly zero) than that in the inner region. The results are shown in Fig.~\ref{tracer-fig} for two different activity strengths. By contrast to what was observed for the activity, the dispersion of the tracers is not subdiffusive. In fact, we have found that the tracers diffuse in the inner region and superdiffuse in the outer region. In the inner active turbulent region, pairs of defects are created and annihilated frequently and move randomly, explaining the diffusive behavior of the tracers. On the other hand, the tracers superdiffusive regime is driven by the motile positive defects (jets), in the outer regions, which propagate almost ballistically as the negative ones are left behind in the inner region. Typical trajectories of the tracers are illustrated in the inset of Fig. 11. It can be seen that they move towards the passive region in the outer region while they move randomly in the inner region. As the motion of tracers is easier to observe experimentally they could be used to detect heterogeneities in the activity field. The observation of the parallel trajectories of tracers, illustrated in the inset, indicates the presence of an active-passive interface. This suggests that the dispersion of particles in the outer regions is faster than that of the scalar activity field. The diffusive and superdiffusive regimes of tracers in active turbulent systems with homogeneous activity have been reported in Ref.~\cite{PhysRevLett.127.118001} based on numerical simulations of a simplified hydrodynamic model, are in line with our observation in the inner and outer regions of the active nematic. The anomalous diffusion, due to Lévy fluctuations, was also observed in experiments with suspensions of algae and bacteria~\cite{10.1098/rsif.2010.0545, Kurtuldu10391, PhysRevLett.84.3017}.

\section{Summary and conclusion}
\label{conclusions-sec}

We investigated the dispersion of activity fields at active-passive interfaces in the active turbulent regime with two aims. The first is the evolution in space and time of an active-passive interface where the local order is quadrupolar. We found that these interfaces become polarized as a result of the different dynamics of the mobile positive (comet-like) and negative (trefoil) 1/2 topological defects, which have an impact on the dispersion of a conserved activity field. We recall that this field results from the product of the concentration of nutrients or other chemicals and their efficiency in producing active stresses in nematics. The second, somewhat more speculative aim, is the investigation of the possible evolution of bacterial shapes to maximize the mixing of nutrients at low Reynolds numbers.

The results were obtained by coupling the advection-diffusion equation with the hydrodynamic equations for active nematics, where the advected solute models the concentration of nutrients or other chemical species capable of producing active stresses in ordered nematics. We found a segregation of positive and negative defects at the active-passive interface, where the mobile positive defects form jets (increasing the dispersion of the activity field) while the negative ones remain close to the interface forming a barrier to the motion of the jets from the active to the passive regions. As a result of this interfacial polarization the dispersion of the activity field is subdiffusive. Not surprisingly, the measured generalized diffusion coefficient increases with the activity strength but somewhat unexpectedly it was found to depend non-monotonically on the aligning parameter. The latter observation suggests that there is an optimal shape to maximize the dispersion of nutrients, which may be relevant to our understanding of the class of biological systems modelled as active nematics.
			

\section*{Acknowledgements}

We acknowledge financial support from the Portuguese Foundation for Science and Technology (FCT) under the contracts: EXPL/FIS-MAC/0406/2021, PTDC/FIS-MAC/28146/2017 (LISBOA-01-0145-FEDER-028146), PTDC/FISMAC/5689/2020, UIDB/00618/2020 and UIDP/00618/2020.

\bibliography{ref} 
\bibliographystyle{rsc} 

\end{document}